\documentclass[12pt,nofootinbib]{revtex4}
\usepackage{amsmath}
\usepackage{amsfonts}
\usepackage{subfigure}
\usepackage{graphicx}
\usepackage{psfrag}
\usepackage{tikz}

\newcommand{\eref}[1]{Eq.~(\ref{#1})}

\newcommand{\nn}{\nonumber}

\newcommand{\fref}[1]{Fig~\ref{#1}}

\def\be{\begin{equation}}
\def\ee{\end{equation}}

\def\bea{\begin{eqnarray}}
\def\eea{\end{eqnarray}}
\def\del{\partial}
\ifx\du\undefined
  \newlength{\du}
\fi
\begin{document}
\title{Superconductivity from D3/D7:\\
 Holographic Pion Superfluid}

\author{Pallab Basu}
 \email{pallab@phas.ubc.ca}

 \author{Jianyang He}
\email{jyhe@phas.ubc.ca}

 \author{Anindya Mukherjee}
 \email{anindya@phas.ubc.ca}

\author{Hsien-Hang Shieh}
\email{shieh@phas.ubc.ca}

 \affiliation{Department of Physics and Astronomy, 
 University of British Columbia,
 6224 Agricultural Road,
 Vancouver, B.C. V6T 1Z1,
 Canada}
\begin{abstract}
 We show that a D3/D7 system (at zero quark mass limit) at finite isospin chemical potential goes through a superconductor (superfluid) like phase transition. This is similar to a flavored superfluid phase studied in QCD literature, where mesonic operators condensate. We have studied the frequency dependent conductivity of the condensate and found a delta function pole in the zero frequency limit. This is an example of superconductivity in a string theory context. Consequently we have found a superfluid/supercurrent type solution and studied the associated phase diagram. The superconducting transition changes from second order to first order at a critical superfluid velocity. We have studied various properties of the superconducting system like superfluid density, energy gap, second sound etc. We investigate the possibility of the isospin chemical potential modifying the embedding of the flavor branes by checking whether the transverse scalars also condense at low temperature. This however does not seem to be the case.
\end{abstract}

\maketitle

\section{Introduction}
The phase structure of QCD (and similar theories) at finite isospin and baryon chemical potential is an interesting issue. At high enough isospin density the color singlet mesonic flavor degrees of freedoms (e.g. pions) may go through a Bose-Einstein condensation. The physical motivation to study such pion superfluid formed at high isospin density is related to the investigation of neutron stars, isospin asymmetric nuclear matter and heavy ion collisions at intermediate energy. Unfortunately this set of problems are difficult to tackle numerically due to the complex nature of the action. Various approaches including lattice simulations are used to investigate the nature of the QCD phase diagram at finite isospin chemical potential and existence of a superfluid like state is argued \cite{Splittorff:2000mm,Son:2000by,Son:2000xc,Toublan:2003tt,Kogut:2002zg,Hao:2006sh}. 

One way to address various facts about gauge theory is to use the gauge gravity duality \cite{Maldacena:1997re} and study a supergravity/tree level string theory to learn about large-N gauge theories. Although such examples do not include QCD or even pure YM theory yet, many qualitatively similar models has been constructed. At the idealized limit where the ratio of flavor and color degrees of freedom is small, one can introduce probe branes in the gravity background to study flavor physics \cite{Karch:2002sh}. In this scenario, the baryon/isospin chemical potential maps to chemical potential for various gauge fields living in the brane. The issue of baryon and isospin chemical potential has been addressed in various type of brane systems \cite{Rozali:2007rx,Mateos:2007vc,Erdmenger:2008yj,Kobayashi:2006sb,Erdmenger:2007ja,Aharony:2007uu,Ghoroku:2007re}. One issue which is relatively less discussed in string theory literature is the issue of flavor superconductivity. There are isospin charged bosonic states with mass of $O(1)$ (e.g pions in QCD), which may be thought as strings which has endpoints on different flavor branes. Such a state may naturally condense as we turn on the isospin chemical potential. Here we discuss such a scenario in a D3/D7 system in the zero quark mass limit. We introduce a couple of branes with no separation between the branes and the branes end on a $AdS_5$ black hole. We turn on a chemical potential corresponding to the $SU(2)$ isospin gauge field living in the world volume of the branes and study the resulting superconducting phase transition and various properties associated with it. 

The plan of the paper is the following. In section \ref{sec:setup} we will set up the probe brane configuration and the equations of motion for the gauge
fields. In sections \ref{sec:phasediag} and \ref{sec:freq}, we establish the superconducting phase transition and study the frequency dependent conductivity which has a pole (corresponds to DC superconductivity) at zero frequency. The speed of second sound is also calculated. Section \ref{sec:statiso} is devoted to DC supercurrent and the phase diagram as a function of the temperature and the velocity of the supercurrent. We will discuss in detail the possibility of other relevant adjoint scalar fields also condense at low temperatures through a similar mechanism in section \ref{sec:what}. In section \ref{sec:conclusions}, we will conclude and point out possible future extensions to the project.

{\bf Note added:} When our work is near completion, another paper \cite{Ammon:2008fc} came which deals with similar questions. These authors have however kept the full DBI action and the resulting details are therefore a little different.

\section{General Setup}
\label{sec:setup}

Let us consider $AdS_5 \times S^5$ in Poincare co-ordinate,
\begin{equation}
 ds^2 = -f(r) dt^2 + \frac{dr^2}{f(r)} + r^2 (dx_1^2 + dx_2^2 + dx_3^2)  + L ^ 2 d\Omega^{ 2}_5
\end{equation}
where
\begin{equation}
f(r) =\frac{r^2}{L^2}-\frac{M}{r^{2}}
\end{equation}
and  $L$ is the radius of the anti-de Sitter space and $M$ is the
the mass of the black hole. In this note we will adopt the
convention that $M =L =1 $. The temperature of the blackhole (and
also the boundary field theory)is given by
\begin{equation}
 T =\frac{1}{\pi}.
\end{equation}
It is more convenient to analyze the system by making a coordinate
transformation $z = 1/r $. The metric becomes:
\begin{equation}
 ds^2 = -f(z) dt^2 + \frac{dz^2}{z^4 f(z)} +  \frac{1} {z^2}  (dx_1^2 + dx_2^2 + dx_3^2)+d\Omega^{2}_5
\end{equation}
with
\begin{equation}
f(z) =\frac{1}{z ^2}-z^2 .
\end{equation}
The horizon is now at $z=1$, while the conformal boundary lives at
$z=0$. In this back ground we will introduce two (or possibly more)
co-incident D7-branes. For simplicity we will consider the zero
quark mass embedding, where the brane is filling the whole $AdS^5$
and wrapping the maximal $S^3$ of $S^5$. In this limit the effective
induced metric on the brane will just be the $AdS_5 \times S^3$
metric.
\begin{equation}
 ds^2_{brane} = -f(z) dt^2 + \frac{dz^2}{z^4 f(z)} +  \frac{1} {z^2}  (dx_1^2 + dx_2^2 + dx_3^2)+ d\Omega_3 ^2
\end{equation}

The effective action for the brane fields is the Born-Infeld action,
\begin{equation}
 S_{DBI}=-T_7 \int d^8 x \sqrt{G +2\pi\alpha ' F}
\end{equation}
In order to consider the system at finite isospin chemical
potential, we will put in a pair of $D7 $ as probe branes. In this
case, $F$ is an $U(2)$ field strength on the world volume of the
probes. We will not focus on baryonic $U(1)_B$ and only investigate terms containing $SU(2)$ isospin gauge fields. The string states which has their endpoints at different D7-branes are charged under the isospin $SU(2)$. The exact form of the non-Abelian DBI action is unknown \cite{Johnson:2000ch}.
To proceed further we will expand the action to leading order in
$\lambda_{YM4} $ keeping only Yang Mills terms. Such simplification has also been employed in studying other aspects of holographic QCD such as the meson
spectra and baryon masses\cite{Sakai:2005yt, Hata:2007mb}. Such an approximation will be more accurate in the limit where the non-Abelian field strengths are small, which may not be the exact situation in our case. However we believe that essential physics is captured. The effective action now takes the form
\begin{equation}
 S_{DBI}=-T_7\frac{(2\pi\alpha ')^ 2}{g_{s}} \int d^8 x\sqrt{-G} TrF^ 2
\propto N_c \int d^8 x\sqrt{-G} TrF^ 2
\end{equation}
where we have scaled out $7+1 $ dimensional Yang Mills coupling
$g_7$ and
\begin{equation}
F^a =\partial A^a +\epsilon^{abc} A^b A^c
\end{equation}

The setup is very similar to that of \cite{Gubser:2008zu}, where the non-Abelian gauge field is shown to condense at low temperature in an AdS black hole background. Due to the non-Abelian nature of the $SU(2)$ symmetry, the $\tau^1$ and $\tau^2$ components of the gauge field are charged under the $\tau^3$ component. Hence turning on a chemical potential for the $\tau^3$ component of the gauge field may lead to a condensation of the other two. This mechanism is very similar to condensation of $U(1)$ charged scalar discussed in \cite{Hartnoll:2008vx}, where a charged black hole may lower its free energy by distributing its charge in a scalar condensate. Turning on a chemical potential for the $\tau^3$ component of the gauge field breaks $SU(2)$ to $U(1)_I$. A condensation of $\tau^1$ or $\tau^2$ component of the gauge field further breaks the $U(1)_I$ symmetry. It should be mentioned that unlike $U(1)_B$ where all charged states are baryonic and has a mass of order $O(N)$, this $U(1)_I$ theory has charged states (mesons) with $O(1)$ mass and their condensation can naturally be studied in terms of the probe brane picture.   

We start with the ansatz
\begin{equation}
 A = A_t \tau^3 dt + B_x \tau^1 dx_1  \\
\label{ansatz}
\end{equation}
We will assume spatial homogeneity in the field theory directions
and our fields will only have dependence on the radial coordinate.
The equations of motion for the fields in this coordinate system
are: \bea
A_t''-\frac{A_t'}{z}-2\frac{B_x^2}{z^2 f} A_t &=&0 \\
B_x''+\left(\frac{f}{f'}+\frac{1}{z}\right) B_x'+\frac{1}{z^2
f}\left( A_t^2 \frac{B_x}{z^2 f} \right)&=&0 \label{maineq} \eea
To require regularity at the horizon we will have to set $A_t=0  $
at $z=1 $. Since we have a set of coupled equations, this will in
turn give the following constraints at the horizon
\begin{eqnarray}
\label{regular}
\nn  B_x '&=& 0 \\
\nn A_t&=&0
\end{eqnarray}
at $z=1$. Examining the behavior of the fields near the boundary,
we find
\begin{eqnarray}
\nn  A_t &\sim &  \mu - \rho z^2 + ...\\
\nn  B_x &\sim & M_x + W_x z^2+ ... \label{asymp}
\end{eqnarray}
Using gauge/gravity duality, $\mu $, $\rho$ are mapped to the isospin 
chemical potential and charge density in the dual field theory,
respectively. $W_x $ is mapped to the expectation value of a meson
operator which condenses at low temperatures. We will set the
non-normalizable mode $M_x $ to zero. In the following we first
establish that the mesonic condensate forms below a critical
temperature.  We compute the time dependent conductivity by turning
on a spatial component for the isospin current as fluctuations.
\begin{equation}
 A_x = X (z) e^{i\omega t}\tau^3 dx_3
\label{isocurrent}
\end{equation}
The equation of motion for $A_x $ is
\begin{equation}
X''+   ( \frac{1}{ z  } + \frac{  f'}{  f }  )X ' - \ B_x^2
\frac{  X  }{ z^2 f } + \frac{ \omega^2 X  }{ z^4 f^2 }=0
\label{Xeq}
\end{equation}
We will choose ingoing boundary condition at the horizon $A_x\propto
(z-1) ^ {-i\omega/4} $.  Asymptotically,
\begin{equation}
  X \sim   S_x  + J_x z^{2} + ...
\label{Xval}
\end{equation}
$J_x $ corresponds to the  isospin current, while $S_x $ gives the
dual current source (superfluid velocity). The conductivity is given by
\begin{equation}
\sigma =Re\left[\frac{J_x}{i\omega S_x}\right]
\label{conduct}
\end{equation}
$X $ can be normalized to one at the horizon. As we will see, the
conductivity has a pole at $\omega = 0 $. This suggests that there
is a DC supercurrent solution.  To find such a solution  we solve
the following set of coupled time independent equations \cite {Basu:2008st,Herzog:2008he}.  Note here the effects of $A_x = X(z)$ on the other components
of the gauge fields are taken  into account, so this is not a fluctuation around the condensate formed by $B_x$. In this case the field $A_x$ has the form $A_x = X (z) \tau^3 dx_3$.
\bea
\label{static_sys}
\nn A_t''-\frac{A_t'}{z}-\frac{B_x^2}{z^2 f} A_t &=&0 \\
B_x''+\left(\frac{f}{f'}+\frac{1}{z}\right) B_x'+\frac{1}{z^2 f}\left( A_t^2 \frac{B_x}{z^2 f}-B_x^3 -X^2 B_x\right)&=&0\\
\nn X''+   ( \frac{1}{ z  } + \frac{  f'}{  f }  )X ' - \ B_x^2\frac{  X  }{ z^2 f }&=&0 \label{current} 
\eea 
The consistency
condition $ A_t = 0 $ at the horizon also imposes the following
constraint on $A_x $  at the horizon
\begin{equation}
  X'= -\ B_x ^2\frac{  X  }{ 4  }
\end{equation}
As we will see shortly, the system \eref{static_sys} reveals a rich phase structure as the boundary value of $A_x$ is tuned.

The convenient physical parameters for us are $(\frac{T}{\mu },\frac{\omega}{\mu }, \frac{{S_x }}{\mu }, \frac{\sqrt[3]{W_x}}{\mu }) $, or$(\frac{T}{\mu }, \frac{\omega}{\mu },\frac{\sqrt[3]{J_x }}{\mu }, \frac{\sqrt[3]{W_x }}{\mu }) $. We will use $ \frac{\sqrt[3]{W_x }}{\mu } $is an order parameter and plot the the phase diagrams as a function of $(\frac{T}{\mu },\frac{{S_x}}{\mu }) $. In practice, we choose to keep the temperature fixed and vary $\mu $ in this paper. The components of gauge fields on the three sphere may also condense through similar mechanism.  We will examine these cases in the appendix. It is also an important question to check whether the isospin chemical potential would modify the embedding of the flavor branes. We will leave the detail of these investigations in a future work.  Here we will begin looking into these questions by checking whether the transverse scalars also condense at low temperature(section \ref{sec:what}). 

\section{Phase Diagram}
\label{sec:phasediag}

\begin{figure}
 \includegraphics[scale=0.5]{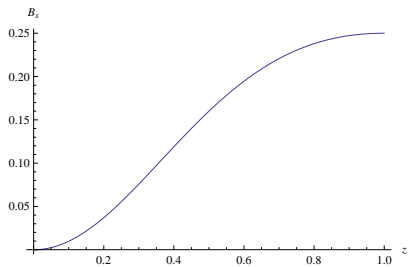}
\caption{Plot of the zero mode at $\mu=4$}
\label{fig:zeromode}
\end{figure}

At high temperature (equivalently small $\mu$) there is only one branch of solution to the set of equation \eref{maineq} given by,
\bea
\label{ordinary}
A_t &=& \mu (1-z^2) \\
\nn B_x &=& 0 
\label{normal}
\eea
This should be interpreted as isospin-charged black hole, where gauge fields are confined in the D7 brane. The dual gauge theory interpretation is a deconfined plasma with non-zero isospin charge. As we increase $\mu$ the effective mass of $B_x$ in \eref{maineq} decreases and $B_x$ develops a zero mode at $\mu = \mu_c = 4$. The existence of this zero mode can be analytically demonstrated. Putting the value of $A_t$ from \eref{ordinary} in the second equation of \eref{maineq} and we get (this small fluctuations analysis does not depend on the any possible cubic terms and hence true for other scalar field ansatzs considered later) ,
\bea
B_x''+\left(\frac{f}{f'}+\frac{1}{z}\right) B_x'+\left( \frac{\mu^2 (1-z^2)^2}{z^4 f^2}\right)B_x=0
\eea
The above equation has an analytic solution for $\mu=4$, given by,
\bea
B_x(z)=\frac{z^2}{(1+z^2)^2}
\eea
The plot of the zero mode is shown in \fref{fig:zeromode},

Any further increment of $\mu$ leads to a condensation of $B_x$. Hence for $1/\mu<0.25$ there is a new branch of solution with non-zero value of $B_x$. Such a solution can be numerically constructed. The associated transition seems to be a second order transition from our numerics. Here we show the plot of condensate (the plot of free energy is provided in section \ref{sec:what}) with $\frac{1}{\mu}$.

\begin{figure}
 \includegraphics[scale=0.6]{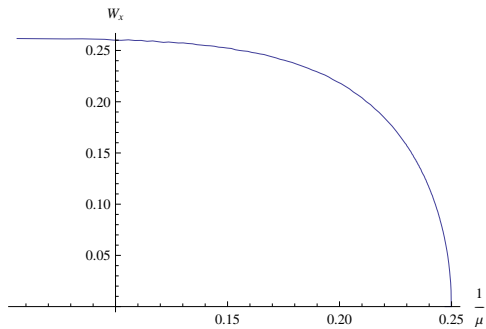}
\caption{Plot of the condensate with $1/\mu$}
\label{fig:condense}
\end{figure}

We find that the condensate levels off at low temperature to $W_x/\mu^3 \approx 0.26$. In terms of more familiar variables critical temperature ($T_c$), the condensate strength can be expressed as $W_x \approx 0.26 4^3 \pi^3 T_c^3 \approx 515.94 T_c^3$ or $W_x ^ \frac{1}{3} \approx 8.01 T_c$.

\subsection{Speed of Second Sound}
\label{sec:2ndsnd}

\begin{figure}
 \includegraphics[scale=1]{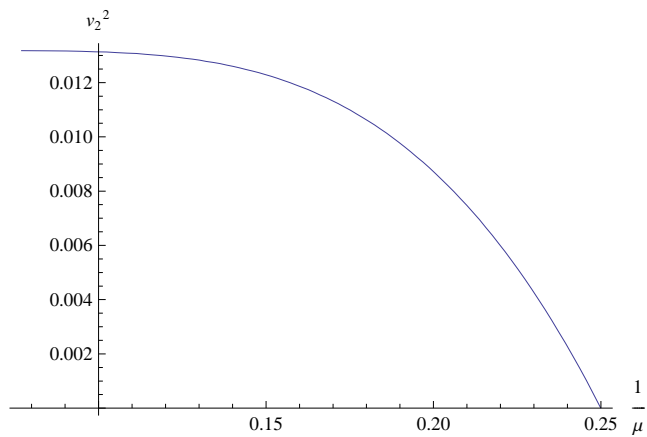}
\caption{Speed of second sound as a function of $1/\mu$}.
\label{fig:2ndsnd}
\end{figure}

The boundary field theory which is dual to the D3/D7 system in $AdS_5 \times S^5$ behaves like a superfluid below the critical temperature. Superfluids are known to exhibit modes known as second sound which are basically temperature waves propagating through the fluid. For a hydrodynamic discussion see \cite{Herzog:2008he}, where this was computed at zero superfluid velocity for the Abelian Higgs model on $AdS_4$. We also compute the speed of second sound in our case. The superfluid velocity now corresponds to $S_x/\mu$, which we set to zero for this computation. The main relation is (see Eq. (18) of \cite{Herzog:2008he}):
\begin{equation}
 \label{2ndsnd}
 v_2^2 = \frac{\frac{\rho_s}{\mu (d-1)}}{\frac{\del^2 P}{\del \mu^2}},
\end{equation}
where $\rho_s$ is the density of the superfluid component and $P$ is the pressure. The pressure can be expressed in terms of the total fluid density $\rho$ by using the equation of state of a perfect fluid $P = \rho/(d-1)$ where $d$ is the dimension in the fluid (boundary) theory\footnote{The perfect fluid approximation is valid here since we are not considering any backreaction due to the metric. Hence there is no viscosity correction which originates from fluctuations of the metric.}. Using this it is fairly simple to compute $v_2^2$. We present the result in Fig \ref{fig:2ndsnd}, where we plot $v_2^2$ as a function of $1/\mu$. At high values of $\mu$ $v_2^2$ approaches a limiting value $v_2^2 = 0.013$.

\section{Frequency Response}
\label{sec:freq}

\begin{figure}
\begin{center}
 \subfigure [Plot of real part of $\frac{\sigma}{\mu}$] {\includegraphics[scale=0.6]{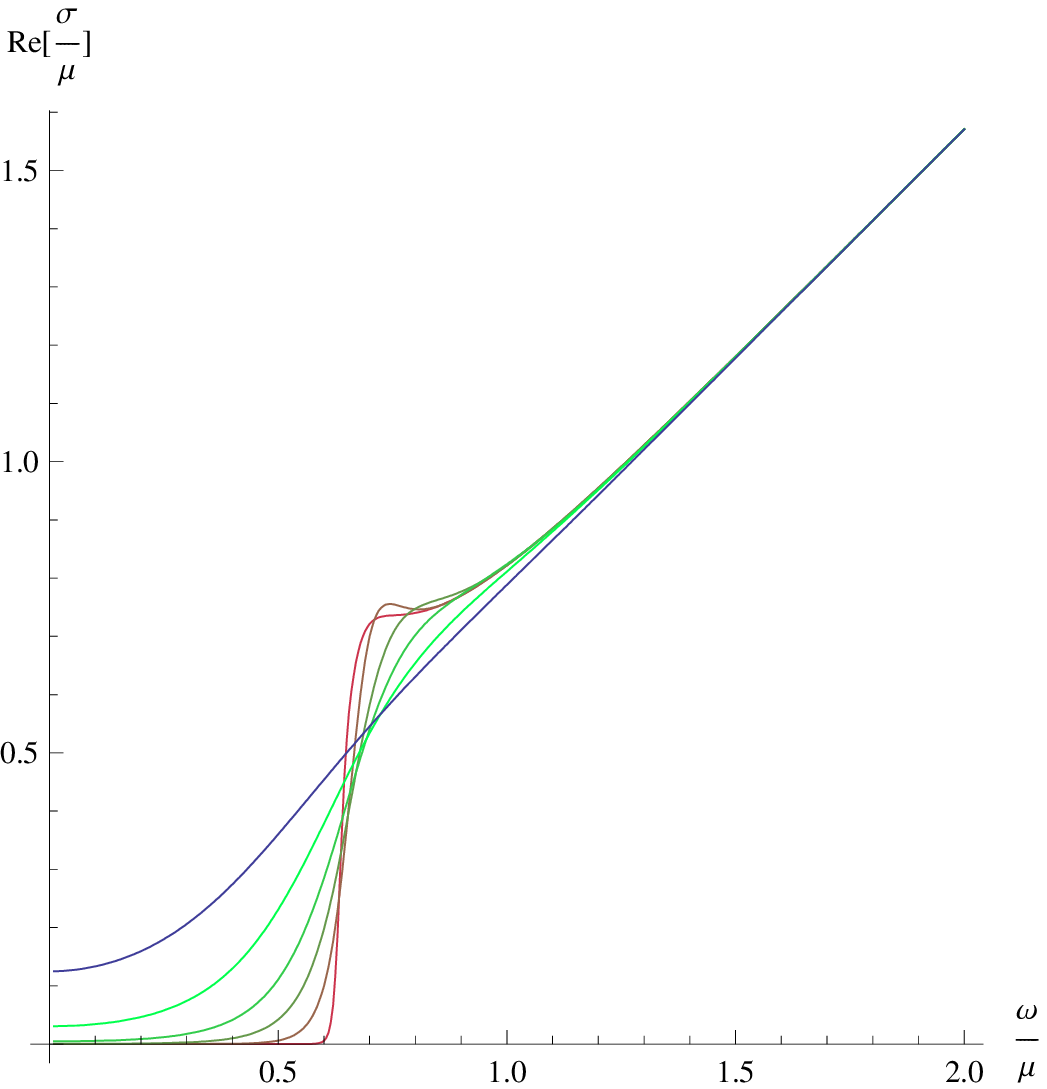}}
 \subfigure [Plot of imaginary part of $\frac{\sigma}{\mu}$] {\includegraphics[scale=0.6]{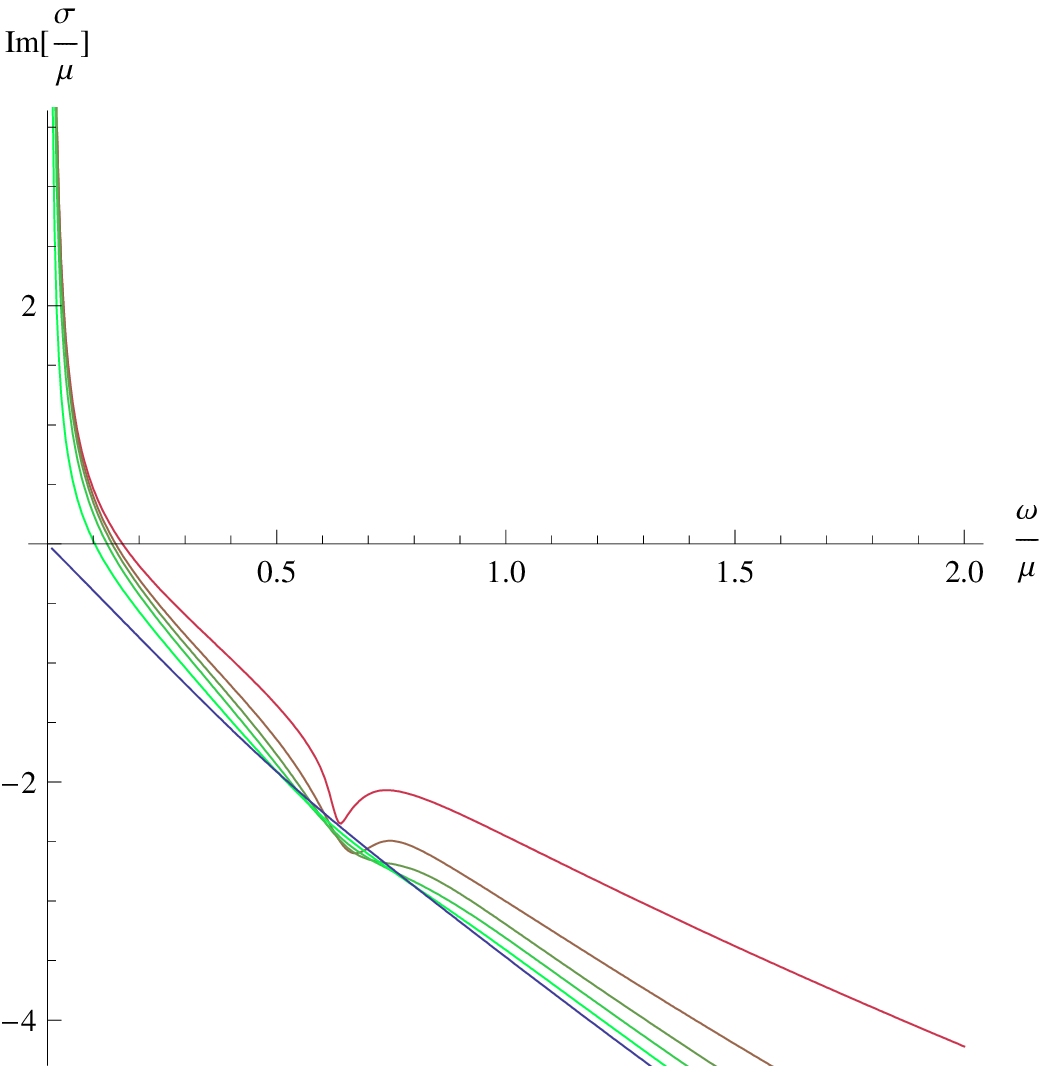}} 
\caption{Plot of the real part of $\frac{\sigma}{\mu}$ with $\mu=7.57 \mu_c, 2.52 \mu_c, 1.71 \mu_c, 1.37 \mu_c, 1.13 \mu_c$ (gradually from red to green curves). The blue curve is for the exact frequency response at $\mu=\mu_c=4$.}
\label{fig:sig}
\end{center}
\end{figure}

\begin{figure}
\begin{center}
 \includegraphics[scale=0.6]{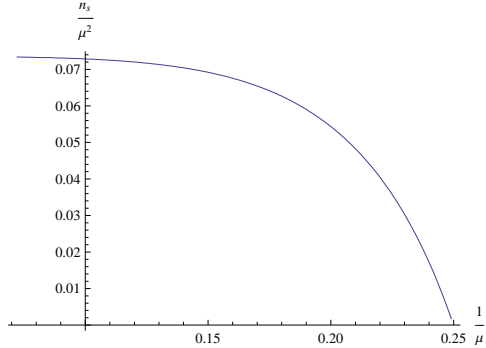}
\caption{Plot of superfluid density with $1/\mu$.}
\label{fig:sufluden}
\end{center}
\end{figure}

Here we will study the frequency dependent conductivity of the spatial component of the isospin current \eref{isocurrent}. Without any condensation of $B_x$ the frequency response can be exactly solved \cite{Horowitz:2008bn}, and is expressed in terms of digamma functions. With the condensation of $B_x$ field, we can still numerically calculate conductivity using \eref{conduct}. We solve the equations \eref{maineq} to get the condensate and on the fixed background given by the solution, we solve \eref{Xeq}. The corresponding $\sigma$ is shown for various values of the parameter $\mu$ in \fref{fig:sig}. 
We find that $Im[\sigma(\omega)] \sim \frac{n_s}{\omega}$ as $\omega \rightarrow 0$. Where $n_s$ is the superfluid density. The plot of the superfluid density is presented in \fref{fig:sufluden}. Near $\mu=\mu_c$, $n_s$ vanishes linearly and is proportional to $\mu-\mu_c$. Fitting a linear function near the critical point we get $n_s \propto 0.1 \mu_c^2 (\mu-\mu_c)$.
A pole at $\omega=0$ implies a $Re[\sigma(0)] \sim \pi n_s \delta(\omega) + terms~ regular~ in~ \omega$. This delta function singularity of the real part of sigma is not captured in the numerics directly. However this corresponds to the superconductivity/superfluidity and consequently we can find a supercurrent/superfluid type of solution (see Sec \ref{sec:statiso}). Unlike \cite{Ammon:2008fc} we do not get any low temperature resonances in the conductivity. Our result is more similar to the zero mass Abelian-Higgs system presented in \cite{Horowitz:2008bn}.\footnote{It seems that in an Abelian-Higgs system in $AdS_5$ resonances occur near conformal mass. \cite{prg}}

As $\int Re[\sigma] d\sigma$ is a temperature invariant quantity, the delta function at $\omega=0$ is compensated by a dip of $Re[\sigma]$ for lower value of frequencies. Also the curves for various temperatures approach the no-condensate curve for higher frequencies.  The dip becomes more prominent as we lower the  temperature (increase $\mu$). It is clear from the diagram that at low temperature (for large $\mu$) $Re[\sigma] \rightarrow 0$. In fact it is expected that at low temperature $Re[\sigma] \sim \exp(-\frac{\Delta_g}{T})$, where $\Delta_g$ is the energy gap of the system. Also looking at the zero temperature limit of the real part of conductivity we see that $Re[\omega]=0$ for $\omega \le \Delta_p$. $\Delta_p$ is similar to the energy of a ``Cooper pair''. The ratio $n_g=\frac{\Delta_p}{\Delta_g}$ gives important information about the nature of the condensate. From our numerics we calculate,
\bea
n_g \approx 1.2 
\eea

\section{Effect of Stationary Isospin Current}
\label{sec:statiso}

\begin{figure}
 \includegraphics[scale=1]{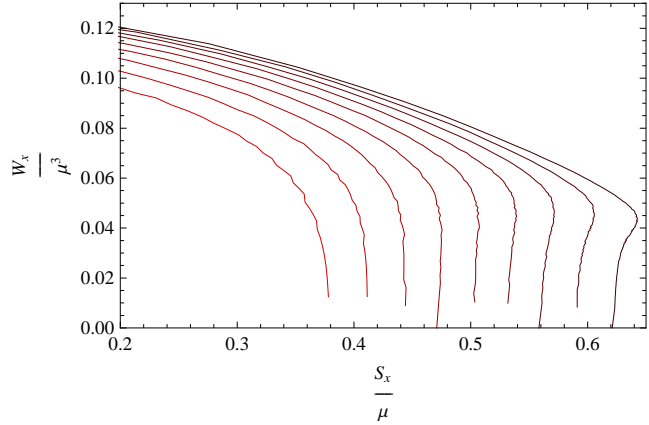}
\caption{Plot of $W_x/\mu^3$ as a function of $S_x/\mu$ at different values of $\mu$. The values of $\mu$ range from 1.24$\mu_c$ to 1.97$\mu_c$ from left to right.}
\label{fig:first2nd}
\end{figure}

In this section we investigate the effects of turning on a finite time-independent isospin $X$ field (recall $A^a_\mu(r) = A_t(r) \tau^3 dt + B_x(r)\tau^1 dx_1 + X(r) \tau^3 dx_3$). We tune the boundary value of $X$, $S_x/\mu$ at a fixed isospin chemical potential $\mu$ to investigate the phase structure. At high enough $\mu$ and in absence of $X$ the gauge component $B_x$ condenses. As we can see from \eref{static_sys}, the effect of $X$ is to increase the effective mass of the $B_x$ field. This will cause the $B_x$ condensate to weaken with increasing $S_x/\mu$. Indeed, this happens, as we can see from Fig \ref{fig:first2nd}. Here we plot the condensate strength as a function of the isospin current source $S_x/\mu$ for different chemical potentials $\mu$. $\mu$ increases from left to right. For strong enough $S_x/\mu$ (above a critical value) there is a phase transition to the normal (non-superfluid) state. The order of this phase transition seems to be $\mu$ dependent. For high $\mu$ (compared to $\mu_c$) the phase transition is first order, i.e. the system discontinuously jumps to the normal state above the critical current velocity $S_x/\mu$. For $\mu$ close to $\mu_c$ the transition becomes second order. The order of the transition changes near $\mu_{sp} = 1.4\mu_c$. Note that in each case the condensate approaches a limiting value at low values of $S_x/\mu$, and this limiting value decreases with decreasing $\mu$.

\begin{figure}
\begin{center}
\subfigure [$\mu = 1.97$] {\label{fig:staila}\includegraphics[scale=0.8]{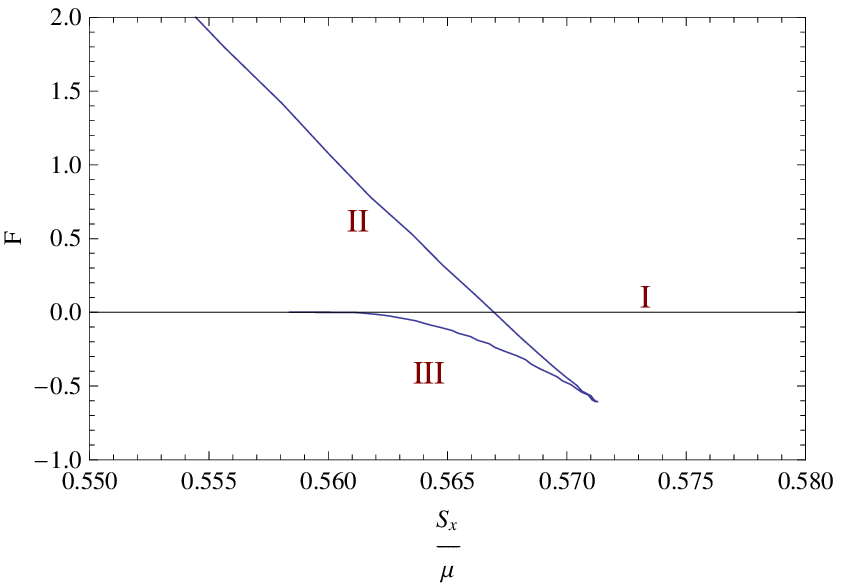}}\hspace{0.5cm}
\subfigure [$\mu = 1.24$]{\label{fig:stailb} \includegraphics[scale=0.75]{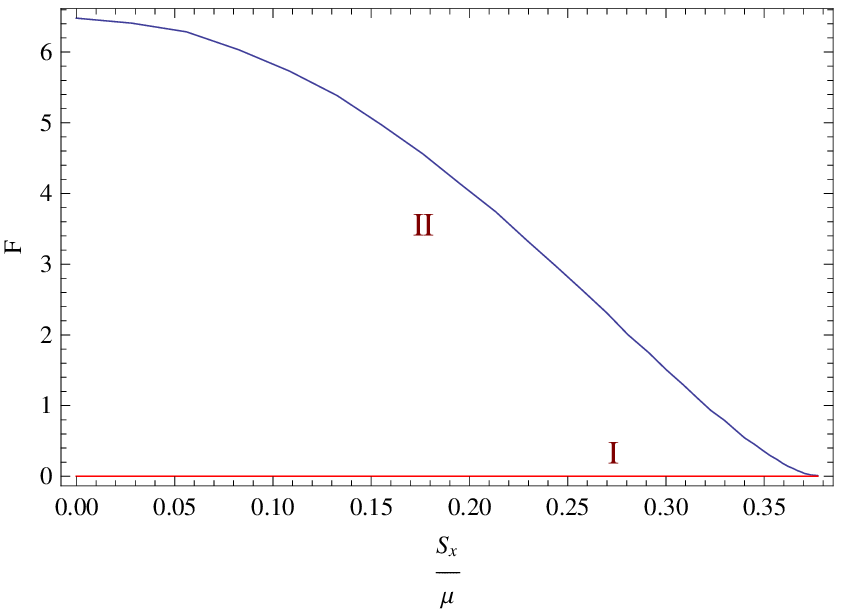}}
\end{center}
\caption{Plot of the difference in free energies of the normal and superconducting branches as a function of $S_x/\mu$}
\label{fig:stail}
\end{figure}

In order to properly see the transition from first to second order, we also plot the difference in free energies of the normal and superconducting branches. In Fig \ref{fig:stail} the left hand plot is for $\mu = 1.97 \mu_c$ while the right hand plot is for $\mu = 1.24 \mu_c$. We see the typical swallow tail shape indicating the abrupt change in dominance from the normal to the superconducting branch at low values of $S_{x}/\mu$. We can understand this curve Fig \ref{fig:staila} in the following manner. As $S_x/\mu$ is lowered from above the critical value $S_{x,c}/\mu$, at first there is only a normal branch (I). At some value $S_{x,N}/\mu = 0.571$ two new branches are nucleated: one of these is stable (II), while the other one is unstable (III). The stable branch starts out with a higher free energy than the normal branch, but as $S_x/\mu$ is lowered further these two branches intersect at $S_{x,c}/\mu = 0.568$, where the branch (II) has the same free energy as branch (I). This is the first order phase transition point below which the system jumps to the superconducting branch (II) which now has lower free energy. At lower values of the chemical potential $\mu$ shown in Fig \ref{fig:stailb} the transition is continuous between branches (I) and (II) at $S_{x,c}/\mu = 0.038$.

\section{What Condenses and What Doesn't} 
\label{sec:what}

There are various type of bosonic fields living in the branes which are charged under $A^3_\mu$ and may condense as we turn on a chemical potential for $A^3_\mu$. These states come from strings ending at different branes. This include the example of gauge fields proportional to $\tau^1$ and $\tau^2$. Together with $A^3$, the trio corresponds to a vector iso-vector operator (flavor current) \footnote{As $m_q=0$ to begin with in our case, there is no stable meson. We only have quasi normal modes corresponding to various brane fields \cite{Kovtun:2005ev}. The dual interpretation of such modes are mesons decaying in gauge theory plasma.} in the boundary theory. We have considered a ansatz like \eref{ansatz}, but more generally one may consider $A = A_t \tau^3 dt + B_x \tau^1 dx_1 + C_x \tau^2 dx_2 $, then action contain a repulsive interaction term ($\propto B_x^2 C_x^2$) between the two, so one may expect that both do not condenses together. And a ansatz of the form $B_x (\tau^1 dx_1 +  \tau^2 dx_2)$ \cite{Gubser:2008zu} will generally have a higher free energy than the ansatz we have chosen. 

\begin{figure}
\begin{center}
 \includegraphics[scale=0.6]{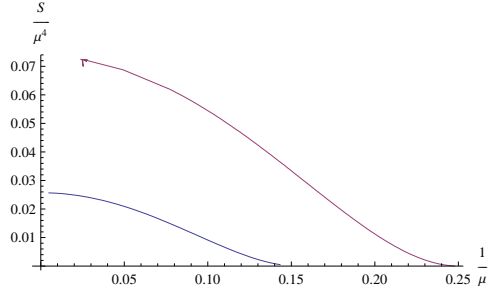}
\caption{Plot of action for phases $B_x \neq 0$ (upper curve) and $\phi \neq 0$ (lower curve).}
\label{fig:action}
\end{center}
\end{figure}

Another possibility is the transverse scalar fields in the brane. Brane fields are invariant under only a $SO(4)\times SO(2)$ subgroup of the full R-symmetry group $SO(6)$ of $S^5$. Two transverse scalar fields correspond to one $SO(2)$-charged iso-vector scalar operator in the boundary theory. The Born-Infeld type effective action for such scalars is given in the appendix. We choose a general ansatz suitable for our case,
\be
\Phi_1=\phi_1\tau^1 , \quad \Phi_2 = \phi_2\tau^2,
\ee
The EOM's  are,
\be
\phi_i''+(-\frac{1}{z}+\frac{f'}{f})\phi_i'+\frac{1}{z^4f}\left(\frac{A_t^2}{f}-\phi_j^2\right)\phi_i=0,
\ee
where $i,j=1,2$ and $i\neq j$  From the discussion of the previous paragraph it is clear that due to the repulsive interaction term, a preferred configuration will have either $\phi_i$'s turned off. One may try to find when such a field becomes unstable in a fixed $A_t$ background. From our numerics we find out that for the solution \eref{normal} such a thing happens at $\mu^s_c \approx 6.57=1.64 \mu_c$ and greater than our $\mu_c=4$ value for $B_x$ field. Hence when we gradually decrease the temperature of our system, $B_x$ condenses before any scalar degree of freedom. One may further ask wheather the resulting superconducting phase with a $B_x$ condensate has a instability for $\phi$ fluctuations. Our numerics answer this question negatively. It seems that some type of ''blocking'' mechanism stops further condensation of more fields. However that does not rule out the possibility of a first order transition between $B_x$ condensed phase and $\phi$ condensed phase. We plot the free energy of both the phases to investigate a possible first order transition \fref{fig:action},
It seems that the $B_x \neq 0$ phase always dominates. However our numerics is not good for the parameter range $\mu >10\mu_c$, also we did not exhaustively searched for the possibility of various mixed phases. We also have to address the issue of curvature of $S^5$ in more detail. The case for gauge fields with $S^5$ indices is also similar to the scalar fields (see appendix), at least for small fluctuations. Hence the blocking mechanism discussed before will work for them too and these fields will not also condensate. However we have not investigated the possible phases and possibility of a first order transition in great detail for these fields.

\section{Conclusions}
\label{sec:conclusions}

In this paper we explored a model which embeds the mechanism of holographic superconductivity/superfluidity in a string theory setup. We have studied a couple of probe D7 branes in $AdS_5 \times S^5$ back ground. We turn on the finite isospin chemical potential (i.e. potential for some of the world volume gauge fields) and have found the existence of a flavored superconducting state at high enough chemical potential. We studied the frequency dependent conductivity and found a delta function pole in the zero frequency limit. This is superconductivity. Consequently we have found a superfluid/supercurrent type solution and studied associated phase diagram. The superconducting transition changes from second order to first order at a critical a superfluid velocity. The holographic dual of such string theory system is large $N_c$, ${\cal N}=4$ supersymmetric gauge theory with $N_f \ll N_c$. In a dual gauge theory such a superconducting state is characterized by mesonic condensates. We have studied various properties of the superconducting system like energy gap, second sound etc. 

Here we have discuss a possibility of a first order transition between various possible condensate. It is also an important question to check whether the isospin chemical potential would modify the embedding of the flavor branes and whether the transverse scalars may condensate. In our case it does not, for some other setup it might. It would be interesting to address such questions in more detail. 

In QCD, the mesonic condensates (pion superfluid) have been argued to exist at finite isospin chemical potential. A natural extension of our work will be to study more realistic holographic models like \cite{Sakai:2005yt,Aharony:2007uu} or in the context of AdS/QCD like theories. It would be interesting to study such phenomenon where the fundamental degrees of freedoms are fermionic. Here we also focused on the zero quark mass case, which may be thought of as a high temperature limit. It would be natural to extend our work to the case of finite quark mass. Another interesting question to investigate is the flavor backreaction \cite{Casero:2006pt,Sin:2007ze,Basu:2008uc}. 
The mesonic operators which condensate in our case are color singlets and do not lead to color superconductivity. It would be interesting to study some models with color superconductivity. 

\section*{Acknowledgements}

We would like to thank Mark van Raamsdonk and Moshe Rozali for helpful discussions and comments on our work. We thank the String Theory Group at UBC for their support and encouragement. PB and AM acknowledge support from the Natural Sciences and Engineering Research Council of Canada.

\section*{Appendix}

Consider D7-branes embedding in $AdS_5\times S^5$ background,
\begin{equation}
d s^2= -f(z)d t^2+\frac{d z^2}{z^4 f(z)} +\frac{1}{z^2}(d x_1^2+d x_2^2 +d x_3^2)
    +d\theta^2+\sin^2\theta\left(d\varphi^2+\sin^2\varphi~ d\Omega_3^2\right),
\end{equation}
the induced metric is
\begin{equation}
d s_7^2= -f(z)d t^2+\frac{d z^2}{z^4 f(z)} +\frac{1}{z^2}(d x_1^2+d x_2^2 +d x_3^2)
    +\sin^2\theta\sin^2\varphi~ d\Omega_3^2.
\end{equation}

According to \cite{Johnson:2000ch}, the leading order DBI action of D$_p$-brane is
\begin{equation}
S_p=-\frac{T_p(2\pi\alpha')^2}{4g_s}\int d^{p+1}\xi~\sqrt{-G_{ind}}{\mathrm{Tr}}
  \left[F_{ab}F^{ab}+2{\cal{D}}_a\Phi_i{\cal{D}}^a\Phi^i+[\Phi_i,\Phi_j][\Phi^i,\Phi^j]\right],
\end{equation}
where scalars $\Phi^i\equiv X^i/(2\pi\alpha')$ are the transverse coordinates, and
covariant derivative is defined as ${\cal{D}}_a\Phi_i=\partial_a\Phi_i+[A_a,\Phi_i]$.

For D$_7$-branes, one has two scalars $\Phi^i$, $i=\theta,\varphi$.

If we turn on the gauge field with ansatz, $A=A_t \tau^3 dt$, where $\tau^a=\tau^a/2i$,
and thus has the commutation relations $[\tau^a,\tau^b]=\epsilon^{abc}\tau^c$.
The scalars take the general form, $\Phi^i=\Phi^i_a\tau^a$. The non-zero components of ${\cal{D}}_a\Phi_i$ are
\be
 {\cal{D}}_t\Phi_i=[A_t\tau^3,\Phi_i]=A_t(\Phi_{i,2}\tau^1-\Phi_{i,1}\tau^2),
 \hspace{1cm} {\cal{D}}_z \Phi_i=\partial_z\Phi_i=(\partial_z\Phi_{i,a})\tau^a.
\ee

The effective action can be written as
\bea
S_7&=&-\frac{T_7(2\pi\alpha')^2}{4g_s}\int d^8\xi~ \frac{\sin^3\theta \sin^3\varphi}{z^5}\Bigg\{ {\mathrm{Tr}}\left(F_{ab}F^{ab}\right)
     \nonumber\\
   && -g_{ii}\left[z^4f(z)\sum_{l=1}^3(\partial_z\Phi_l^i)^2-f(z)A_t^2\left((\Phi_1^i)^2+(\Phi_2^i)^2\right)\right]
       -\frac{1}{2}\sin^2\theta \left[\sum_{l\neq m}^{1,2,3}(\Phi_l^{\theta}\Phi_m^{\varphi}-\Phi_m^{\theta}\Phi_l^{\varphi})^2\right] \Bigg\},
       \nonumber\\
\eea
where $g_{ii}$ are 10-dimensional metric with $i=\theta$ or $\varphi$.

Introduce $\Phi_{\pm}^i\equiv (\Phi_1^i\pm i\Phi_2^i)/2$, we find the equations of motion for $(\Phi_3^i,\Phi_{\pm}^i)$,
\bea
g_{ii}\Bigg[\frac{f(z)}{z}\partial_z^2\Phi_3^i &+& \left(-\frac{f}{z^2}+\frac{f'(z)}{z}\right)\partial_z\Phi_3^i\Bigg]
 = 2\frac{\sin^2\theta}{z^5}
    \Bigg[\Phi_3^i((\Phi_-^j)^2+\Phi_+^j\Phi_-^j)-\Phi_3^j(\Phi_+^i\Phi_-^j +\Phi_-^i\Phi_+^j)\Bigg], \nonumber\\
g_{ii}\Bigg[\frac{f(z)}{z}\partial_z^2\Phi_+^i &+& \left(-\frac{f}{z^2}+\frac{f'(z)}{z}\right)\partial_z\Phi_+^i\Bigg] \nonumber\\
 &=& \frac{1}{z^5}\Bigg\{ -\frac{g_{ii}}{f}A_t^2\Phi_+^i
    +\sin^2\theta\Big[ \Phi_+^i~\Phi_+^j\Phi_-^j+\Phi_-^i(-(\Phi_+^j)^2+(\Phi_3^j)^2)-\Phi_3^i~\Phi_-^j\Phi_3^j\Big]\Bigg\} \nonumber\\
g_{ii}\Bigg[\frac{f(z)}{z}\partial_z^2\Phi_-^i &+& \left(-\frac{f}{z^2}+\frac{f'(z)}{z}\right)\partial_z\Phi_-^i\Bigg] \nonumber\\
 &=& \frac{1}{z^5}\Bigg\{ -\frac{g_{ii}}{f}A_t^2\Phi_-^i
    +\sin^2\theta\Big[ \Phi_-^i~\Phi_+^j\Phi_-^j+\Phi_+^i(-(\Phi_-^j)^2+(\Phi_3^j)^2)-\Phi_3^i~\Phi_+^j\Phi_3^j\Big]\Bigg\},
\eea
where $i,j\in (\theta,\varphi)$ and $i\neq j$.

Consider fluctuations, one can set $\sin\theta=1$, and thus $g_{ii}=1$ in above equations. If we restrain our discussion in the ansatz
\be
\Phi_1=\phi_1\tau^1 , \quad \Phi_2 = \phi_2\tau^2,
\ee
the EOM's can be simplified,
\be
\phi_i''+(-\frac{1}{z}+\frac{f'}{f})\phi_i'+\frac{1}{z^4f}\left(\frac{A_t^2}{f}-\phi_j^2\right)\phi_i=0,
\ee
where $i,j=1,2$ and $i\neq j$.

For the simplest case, with $\Phi=\phi(z)(\tau^1d x^1+\tau^2dx^2)$, the equation of motion is
\be
\phi''+(-\frac{1}{z}+\frac{f'}{f})\phi'+\frac{1}{z^4f}(\frac{A_t^2}{f}\phi-\phi^3)=0.
\ee
If the gauge fields on the 3-sphere instead of the position scalars of 
D7-branes are turned on, one would find similar condensation.
In an ansatz
\be
A\sim A_t(z)\tau^3 d t +A_{\theta}(z)\tau^1 d\theta,
\ee
the equations of motion are
\bea
A_t''-\frac{1}{z}A_t'-\frac{A_\theta^2}{z^4f}A_t=0, \nonumber\\
A_{\theta}''+\left(-\frac{1}{z}+\frac{f'}{f}\right) 
A_{\theta}'+\frac{A_t^2}{z^4f^2}A_{\theta}=0.
\eea
Or in another ansatz,
\be
A\sim A_t\tau^3 dt+A_{\theta}(\tau^1 d\theta+\tau^2\sin\theta d\varphi),
\ee
where $(\theta,\varphi)$ are angle coordinates on 3-sphere of D7-brane 
worldvolume $AdS_5\times S^3$.
The EOM's turn to be
\bea
A_t''-\frac{1}{z}A_t'-\frac{2A_{\theta}^2}{z^4f}A_t=0, \nonumber \\
A_{\theta}''+\left(-\frac{1}{z}+\frac{f'}{f}\right)A_{\theta}'+\frac{1}{z^4f}\left(\frac{A_t^2}{f}A_{\theta}-A_{\theta}^3\right)=0.
\eea

\bibliography{pion.bib}
\end{document}